\documentclass[aps,twocolumn,showkeys,groupedaddress,amsmath,nofootinbib,amssymb,showpacs,floatfix]{revtex4-1}
\usepackage{graphicx}  
\usepackage{dcolumn}   
\usepackage{color}
\usepackage{psfrag}
\usepackage{bbm,bm}
\usepackage{epsfig}
\usepackage{hyperref}
\allowdisplaybreaks
\begin{document}
\def\be{\begin{equation}}
\def\ee{\end{equation}}
\def\bea{\begin{eqnarray}}
\def\eea{\end{eqnarray}}
\def\ba{\begin{array}}
\def\ea{\end{array}}
\def\ben{\begin{enumerate}}
\def\een{\end{enumerate}}
\def\nab{\bigtriangledown}
\def\tpi{\tilde\Phi}
\def\nnu{\nonumber}
\newcommand{\eqn}[1]{(\ref{#1})}
\def\bw{\begin{widetext}}
\def\ew{\end{widetext}}
\newcommand{\half}{{\frac{1}{2}}}
\newcommand{\vs}[1]{\vspace{#1 mm}}
\newcommand{\dsl}{\pa \kern-0.5em /} 
\def\a{\alpha}
\def\b{\beta}
\def\g{\gamma}\def\G{\Gamma}
\def\d{\delta}\def\D{\Delta}
\def\ep{\epsilon}
\def\et{\eta}
\def\z{\zeta}
\def\t{\theta}\def\T{\Theta}
\def\l{\lambda}\def\L{\Lambda}
\def\m{\mu}
\def\f{\phi}\def\F{\Phi}
\def\n{\nu}
\def\p{\psi}\def\P{\Psi}
\def\r{\rho}
\def\s{\sigma}\def\S{\Sigma}
\def\ta{\tau}
\def\x{\chi}
\def\o{\omega}\def\O{\Omega}
\def\k{u}
\def\pa {\partial}
\def\ov{\over}
\def\nn{\nonumber\\}
\def\ud{\underline}
\def\ct{\textcolor{red}{\it cite }}
\def\qq{\text{$Q$-$\bar{Q}$ }}



\title{\large{\bf Confinement and Pseudoscalar Glueball Spectrum from Anisotropic Non-susy D$2$ Brane under Hawking-Page transition}}
\author{Adrita Chakraborty}
\email{adimanta09@iitkgp.ac.in}
\affiliation{Department of Physics,\\
   Indian Institute of Technology Kharagpur,\\
   Kharagpur 721302, India}
\author{Kuntal Nayek}
\email{kuntal.nayek@iitkgp.ac.in}
\affiliation{Department of Physics,\\
   Indian Institute of Technology Kharagpur,\\
   Kharagpur 721302, India}
\date{\today}

\begin{abstract}
   Here we analyse the low energy confining phase of the $2+1$ dimensional quenched QCD-like theory in anisotropic non-supersymmetric D$2$ brane background with the holographic approach. Two key features of QCD -- flux-tube tension and mass spectrum of pseudoscalar glueball are studied. The non-supersymmetric D$2$ branes with anisotropy in time direction are considered as the gravity theory. Tuning the anisotropic parameter, we get the Hawking-Page transition in the gravity background. On the dual theory, this refers the confinement-deconfinement phase transition. Using the Wilson loop, the linear confinement at this regime is effectively confirmed by calculating the flux tube tension $\sigma$ from the Nambu-Goto action of a test string with its endpoints located at the boundary. In the next part, we have illustrated numerically the mass spectrum $M_{0^{-+}}$ of the pseudoscalar glueball states. We have found, at the transition point, that the glueball mass becomes trivially small. Finally we relate the anisotropy parameter with the temperature and show the variation of the glueball mass at various temperature. Using the Lattice QCD data of $\sqrt{\sigma}$, it is found that $M_{0^{-+}}=2.3$GeV and $M_{0^{-+}}^*=3.8$GeV at $T=0$ whereas $M_{0^{-+}}=0.9$GeV and $M_{0^{-+}}^*=1.1$GeV at $T=T_c$. 
\end{abstract}


\maketitle

\tableofcontents

\section{Introduction}
Since last few decades, there are substantial research going on to establish the perturbative QCD from both the theoretical and experimental stances. On the contrary, the non-perturbative strongly coupled regime still remains as one of the most intriguing quests due to the lack of vigorous understanding of the non-perturbative regime of QFT's. However, some theoretical models namely, chiral perturbation theory, heavy baryon theory, bag model, Skyrme model have put forward plausible observations on the non-perturbative QCD but those are not sufficient for a complete picture of this regime. On the other hand the lattice gauge theory has been proved to be an useful theoretical tool. But it lacks real-time dynamics as it deals with the Euclidean signature. Again, the finiteness of the lattice makes the continuum limit much difficult to compute. In this context, the celebrated AdS/CFT \cite{Maldacena:1997re,Aharony:1999ti} duality in string theory is found to be enough promising to handle the non-perturbative gauge theory. According to this, the operators in the strongly coupled gauge theory correspond to the fields in the equivalent gravity theory. Therefore the value of the observable remains unchanged. So, nowadays the holography is a popular method to study the non-perturbative QCD from the perturbative gravity theory.

The AdS/CFT correspondence was originally defined for the supersymmetric extremal D$3$ brane \cite{Maldacena:1997re,Aharony:1999ti}. It says that the $3+1$ dimensional pure super-Yang-Mills theory is holographically equivalent to the gravity theory on a $4+1$ dimensional pure Anti de Sitter geometry. However it is formulated for the conformal theories, the duality can be nicely extended for the non-supersymmetric, non-conformal theories \cite{Nayek:2015tta,Nayek:2016hsi} without breaking the holographic conditions. As we have seen, besides the BPS D$p$ brane, there is a set of nonsupersymmetric (non-susy) D$p$ brane solutions of type-II supergravity \cite{Lu:2004ms}. These brane solutions also follow the gauge/gravity correspondence and gives a QCD-like, non-conformal gauge theory as the dual theory. Like the AdS/CFT correspondence, in planner limit, we choose the number of non-susy D brane to be very large ($N_c\gg 1$) and the gravity theory to be perturbative ($g_s\ll 1$). The Yang-Mills coupling ($g_\text{YM}^2N_c\sim g_s N_c$) of the dual non-conformal gauge theory is large and finite, i.e., the gauge theory is non-perturbative. Because of the non-conformality, in this case, the gauge theory experiences a fixed mass scale which can be interpreted as the Landau fixed point $\Lambda$ of the QCD-like theory \cite{Nayek:2016hsi}. Again for the dilatonic-gravity theory, the dilaton field represents a non-constant effective gauge coupling which varies with the energy scale of the theory. Therefore the non-supersymmetric gauge/gravity duality has quite efficiently estimated salient QCD-like behaviours like the low energy confinement, gluon condensate, effective running coupling, chiral symmetry breaking etc \cite{Witten:1998zw,Constable:1999ch,Ooguri:1998hq,Csaki:1998qr,Babington:2003vm,Csaki:2006ji,Kim:2007qk,Polchinski:2002jw} both in the three and four dimensions as well as some properties of Quark-Gluon Plasma \cite{Chakraborty:2017wdh}. Since the dual gauge theory is strongly coupled, according to holography, it does not allow the asymptotic freedom \cite{Polchinski:2001tt} in this set-up and we can not find its the perturbative regime. Therefore the gauge/gravity duality in non-susy D brane gives a non-perturbative QCD-like theory. However there are some other approaches to study the QCD-like theory with holography \cite{Witten:1998zw}. One can use circular compactification on one of the longitudinal directions of the BPS D$3$ or D$4$ brane to study three or four dimensional QCD respectively. This compactified direction breaks the conformal symmetry and gives a mass scale in the gauge theory. Also there are some Lattice QCD calculations where people have studied various properties of the non-perturbative QCD \cite{Morningstar:1999rf,Teper:1997tq,Miller:2006hr,Lucini:2001ej,Brandt:2017yzw}. 

The heavy ion collisions at RHIC, LHC has unraveled the quark-gluon plasma (QGP) as a strongly interacting relativistic fluid at high energy. At that plasma state, the hadrons completely dissolve into free quarks and gluons which is called the deconfined state of QCD. Whereas at the low energy, the gluons form bound states -- glueballs, which is named as the confined phase. As the temperature decreases the deconfined phase starts to proceed towards the confinement due to the dominating self interaction, and gives a phase transition at a finite critical temperature (expected to be $150-200$MeV). However according to some recent works \cite{Mandal:2011ws,Bergner:2021goh}, this transition is not a sharp-edged transition rather a cross-over between these two phases. So, in a range of temperature around the transition point it is expected to have a mixed state of gluons and glueballs -- named as partially deconfined state. Since the glueball of the confined state completely annihilates into the free gluons in the deconfined state, the mass spectrum of the glueball is expected to show some significant behaviour during the transition. In this article we are intended to present such behaviour of the pseudoscalar glueball along with the QCD string tension in a $2+1$ dimensional QCD-like model.
 
In this article, to study the $2+1$ dimensional QCD-like theory, we consider a stack of $N_c$ number of dilatonic, non-susy, anisotropic D$2$ brane. Here the anisotropy has been introduced along the time direction of the $2+1$ dimensional worldvolume. The anisotropic D brane solutions involve an extra parameter than the isotropic one, which controls the anisotropy and can be called as the anisotropy parameter. Now varying this parameter one can arrive the Hawking-Page (HP) transition for the non-susy D$2$ brane. Initially the decoupled geometry of the aforementioned gravity theory is anisotropic non-AdS but assymptotically it gives a thermal AdS geometry in $3+1$ dimension. At a particular value of the anisotropy parameter, the gravity background transformed into the black D$2$ brane \cite{Horowitz:1991cd}. This transformation of the gravity theory represents the confinement-deconfinement transition in the $2+1$ dimensional gauge theory. Therefore to study the phase transition in this $2+1$ dimensional QCD-like theory, we study the confinement and pseudoscalar glueball spectrum in this anisotropic gravity background. Then we analyse the evolution of those quantities at the HP transition. Here we take an open string as a probe and a $Q-\bar{Q}$ pair on the boundary. The open string hanging into the bulk, connects these $Q$ and $\bar{Q}$ at its end points. Now we employed the holographic conditions which says that the thermal expectation value of the Wilson loop \cite{Maldacena:1998im,Rey:1998ik,Rey:1998bq,Brandhuber:1998er,Brandhuber:1998bs} is the minimal worldsheet area swept out by the open string in the bulk theory. It computes the $Q-\bar{Q}$ binding potential. Further, taking the $Q-\bar{Q}$ separation large enough we find the linear potential and the flux tube tension. As the anisotropy parameter is zero, we can recover the flux tube tension of the isotropic background \cite{Chakraborty:2020sty}. In the next part we take the axion field's fluctuation in this gravity background. The axion field of the bulk couples to the pseudoscalar glueball on the boundary. So we calculate the pseudoscalar glueball mass from the equation of motion of the axion's fluctuation using the WKB approximation. However there are various articles on the $2+1$ dimensional Yang-Mills theory where the pseudoscalar glueball spectrum has been found analytically \cite{Karabali:1995ps,Karabali:1996je,Karabali:1996iu,Karabali:1998yq,Nair:2002yg} as well as using the holographic \cite{Aharony:1999ti,Csaki:1998qr,Hong:2010sb} and Lattice QCD \cite{Teper:1998te,Teper:1993gm,Philipsen:1996af} technique. Here we show the variation of the spectrum during the phase transition. We also plot the ratio of the glueball mass to the square root of the string tension and it remains constant. In our observation, we find that the glueball mass decreases significantly near the transition point and becomes small but non-zero at the transition point. However the mass-gap between two consecutive energy levels becomes trivially small. The existence of this non-vanishing glueball mass at the transition point can be explained in terms of the partial deconfinement. Finally we give an empirical interpretation of temperature in terms of the anisotropy parameter of the gravity background. This relation is argued by two facts; (i) the decoupled gravity background is asymptotically AdS, and (ii) it shows HP transition.  

Our paper detail the study in the following sections as mentioned below. In section II the non-supersymmetric anisotropic D2 brane solution has been revisited followed by its decoupled geometry at the low energy limit in section III. Section IV is devoted to establish the confinement property of our theory. Computation of pseudoscalar glueball masses has been covered in section V which is then followed by discussions and concluding remarks along with future directions.

\section{Non-susy anisotropic D$2$ brane}
The family of non-supersymmetric Dp brane solutions of type-II supergravity includes some of its sector showing anistropic behaviour in the directions of worldvolume \cite{Lu:2004ms,Lu:2007bu}. As we have seen, due to unequal ADM mass and charge, the non-susy isotropic branes do not support the well-known BPS bounds and hence cause the breaking of conformal supersymmetry of the gauge theory living in their respective worldvolume. In the extremal limit, they are found to merge into the BPS ones. Similar nature is exhibited by the gauge theories associated with the anisotropic sector. However, these solutions involve one extra parameter than the isotropic branes, which comes into play because of the anisotropy along one of the worldvolume directions -- say, anisotropy parameter. At a particular value of this anisotropy parameter, the anisotropic non-susy D$2$ brane reduces to the black D$2$ brane. Therefore varying that parameter we get the Hawking-Page transition and that parameter can be related to a temperature like quantity in the gauge theory \cite{Nayek:2021ded}. Like the isotropic non-susy brane, other parameters of the gravity theory are related to the gluon condensate and the fixed mass scale of the corresponding gauge theory \cite{Nayek:2016hsi,Chakraborty:2020sty}. Following the equation of motion of the gravity background, all of these parameters are equipped with suitable parametric relations. Akin to the case for isotropic non-susy branes, the low energy decoupled geometry of the anisotropic brane is not conformally invariant. The non-conformality of this D$2$ brane reasons the presence of a fixed energy scale. Again according to the holographic dictionary, the non-constant dilaton field claims the effective coupling of the corresponding gauge theory to be energy-scale dependent. Here, the anisotropic non-susy D$2$ brane can be written in Einstein frame from the eq. (4) $\&$ (5)  of \cite{Lu:2007bu} by putting $p=2$ and $q=0$ as
\begin{eqnarray}
ds^2 & = & F^\frac{3}{8}\left(H\tilde H\right)^\frac{2}{5}\left(\frac{H}{\tilde H}\right)^{\frac{\d_1}{4}+\frac{1}{10}\d_2+\frac{1}{10}\d_0}\left(dr^2+r^2d\Omega_6^2\right)\nn
&& +F^{-\frac{5}{8}}\left(-\left(\frac{H}{\tilde H}\right)^{\frac{\d_1}{4}+\half\d_2+\half\d_0}dt^2+\right.\nn
&& \left.\left(\frac{H}{\tilde H}\right)^{-\frac{3}{4}\d_1-\frac{3}{2}\d_2+\half\delta_0}dx_1^2+\left(\frac{H}{\tilde H}\right)^{-\frac{3}{4}\d_1+\half\d_2-\frac{3}{2}\d_0}dx_2^2\right)\nn
e^{2\phi} &=& g_s^2F^\half\left(\frac{H}{\tilde H}\right)^{\d_1-2\d_2-2\d_0}\nn
F_{[6]} & = & \hat{Q}\text{Vol}(\Omega_6)
\end{eqnarray}
where the harmonic functions are 
\begin{eqnarray}
F & = & \left(\frac{H}{\tilde H}\right)^\alpha\cosh^2\theta-\left(\frac{H}{\tilde H}\right)^{-\beta}\sinh^2\theta\\
H & = & 1+\frac{\omega^5}{r^5},~~
\tilde H = 1-\frac{\omega^5}{r^5}
\end{eqnarray}
Here the $2+1$-dimensional world volume is defined with the coordinates $\left(t, x_1, x_2\right)$, whereas the seven-dimensional transverse space is defined by the spherical coordinates $\left(r, \Omega_6\right)$. $r$ being the radial coordinate defines the energy scale of the worldvolume theory. The bulk geometry experiences a singularity at $r=\omega$ and asymptotically becomes $9+1$-dimensional Minkowski metric as $r\to\infty$. The dilaton field $\phi$ is non-constant and function of $r$ in this background and asymptotically reduces to $\phi_0$. Thereby it is obvious that the effective coupling ($\sim e^\phi$) is also a function of the energy scale. $\theta$ is the dimensionless charge parameter, related to the total Ramond-Ramond (RR) charge Q of the brane. The parameters follow the mutual relation between them as
\begin{eqnarray}
&& \alpha-\beta=-\frac{3}{2}\delta_1\nn
&& \half \d_1^2+\half\alpha\beta+\frac{2}{5}\d_2\d_0=\frac{6}{5}(1-\d_2^2-\d_0^2)\nn
&& \hat{Q}=5(\a+\b)\omega^5\sinh2\theta
\end{eqnarray}
Now we take a convenient choice for $\delta_2$ as $\delta_2=\d_0$ and then replace $\d_0$ using the relation $\d_1+2\d_0=2\d$ to ensure of the presence of anisotropy only along the time direction of the worldvolume. This modifies the metric as well as the associated dilaton and flux as
\begin{eqnarray}
ds^2 & = & F^\frac{3}{8}\left(H\tilde H\right)^\frac{2}{5}\left(\frac{H}{\tilde H}\right)^{\frac{3}{20}\d_1+\frac{\d}{5}}\left(dr^2+r^2d\Omega_6^2\right)\nonumber\\
&& +F^{-\frac{5}{8}}\left(\frac{H}{\tilde H}\right)^{-\frac{1}{4}\d_1-\d}\left(-\left(\frac{H}{\tilde H}\right)^{2\d} dt^2+dx_1^2+dx_2^2\right)\nn
e^{2\phi} &=& g_s^2F^\half\left(\frac{H}{\tilde H}\right)^{3\d_1-4\d}\nn
F_{[6]} & = & \hat{Q}\text{Vol}(\Omega_6)
\label{metric1}
\end{eqnarray}
The second of three parametric relations reduces to
\begin{equation}
 \d_1^2+\frac{5}{12}\alpha\beta+\frac{7}{3}(\d^2-\d\d_1)=1,
\label{parameter}
\end{equation}
whereas the other two remain the same. We consider the coordinate transformation for radial coordinate $r$ as
\be
r=\rho\left(\frac{1+\sqrt{G(\rho)}}{2}\right)^\frac{2}{5} \quad\text{where, }G(\rho)=1+\frac{\rho_2^5}{\rho^5}
\label{HF1}
\ee 
and $\rho_2^5=4\omega^5$. This reduces the harmonic function $F$ as 
\be\label{HF2}
F = G^\frac{\a}{2}\cosh^2\theta-G^{-\frac{\b}{2}}\sinh^2\theta
\ee
Now the new coordinate $\rho$ defines the relevant length scale of the theory herein and $\rho_2$ is a constant fixed point on that length scale having the mass dimension $-1$. Following the BPS brane's terminology, we can call $\rho_2$ as the mass parameter. In this new coordinate the singularity of the background sits at $\rho=0$.
Now the definition of the harmonic function $G(\rho)$ demands its value to be always greater than $1$. Consequently, the validity of the metric in the range $0 < \rho < \infty$ refers $F(\rho)$ to be positive assuming $\a>\b$. In this new radial coordinate $\rho$, the metric \eqref{metric1} becomes
\begin{eqnarray}
ds^2 & = & F^\frac{3}{8}G(\rho)^{\frac{1}{5}+\frac{3}{40}\d_1+\frac{\d}{10}}\left(\frac{d\rho^2}{G(\rho)}+\rho^2d\Omega_6^2\right)\nonumber\\
&& +F^{-\frac{5}{8}}G(\rho)^{-\frac{1}{8}\d_1-\frac{1}{2}\d}\left(-G(\rho)^\d dt^2+dx_1^2+dx_2^2\right)\nn
e^{2\phi} &=& g_s^2F^\half G(\rho)^{\frac{3}{2}\d_1-2\d}\nn
F_{[6]} & = & \hat{Q}\text{Vol}(\Omega_6)
\label{metric2}
\end{eqnarray}
 Without loosing any information, to simplify our computation, we use a particular constraint on the parameters by choosing $\a+\b=2$. This gives $\a = 1-\frac{3}{4}\d_1$ and $\b = 1+\frac{3}{4}\d_1$. With these values of $\a$ and $\b$ in hand, we reduce the expression of $F(\rho)$ as $F(\rho)=G^{-\frac{\b}{2}}f(\rho)$, with\\ $f(\rho)=1+\frac{\rho_2^5}{\rho^5}\cosh^2\theta$. We can also solve the parametric equation (\ref{parameter}) and write $\d_1$ as a function of $\d$ as follows.
\bea
\d_1=\frac{4}{21}\left(8\d\pm\sqrt{21-20\d^2}\right)\nonumber
\eea
This defines the range of $\d$ as $-\sqrt{\frac{21}{20}}\leq\d\leq\sqrt{\frac{21}{20}}$ for real $\delta_1$. Among these two roots of $\d_1$, we will use the root with `$+$' sign in our numerical studies. Because of this choice we shall have a dilaton field which decreases monotonically with $|\d|$ in the range $-\sqrt{\frac{21}{20}}\leq\d\leq 0$ and, in addition, this indicates the QCD like properties of the effective gauge coupling. Therefore, considering all of the parametric relations it is quite evident that we are left with three independent parameters, namely $\rho_2,\,\theta$ and $\delta$. With this constraint relation of the parameters, the metric \eqref{metric2} in string frame can be written as,
\begin{eqnarray}
ds^2 & = & f^\frac{1}{2}G(\rho)^{-\frac{1}{20}+\frac{21}{80}\d_1-\frac{2}{5}\d}\left(\frac{d\rho^2}{G(\rho)}+\rho^2d\Omega_6^2\right)\nonumber\\
&& +f^{-\frac{1}{2}}G(\rho)^{\frac{1}{4}+\frac{7}{16}\d_1-\d}\left(-G(\rho)^\d dt^2+dx_1^2+dx_2^2\right)\nn
e^{2\phi} &=& g_s^2f^\half G(\rho)^{-\frac{1}{4}+\frac{21}{16}\d_1-2\d}\nn
F_{[6]} & = & \hat{Q}\text{Vol}(\Omega_6)
\label{Stringframe}
\end{eqnarray}

The Hawking-Page transition \cite{Hawking:1982dh} in the anisotropic non-susy D$p$ brane background has been already discussed in \cite{Lu:2007bu}. Here we briefly present the same for the above gravity background. At $\d=-1$ we have $\d_1=-\frac{4}{3}$ and $-\frac{12}{7}$. Then we shift the singularity of the metric by taking a coordinate transformation $r^5=\rho^5+\rho_2^5$. For $\d=-1$ and $\d_1=-\frac{12}{7}$, the metric \eqref{Stringframe} reduces to the standard black D$2$ brane in the Einstein frame of the following form.
\begin{eqnarray}
ds^2 & = & \bar H(r)^\frac{3}{8}\left(\frac{dr^2}{\tilde f(r)}+r^2d\Omega_6^2\right)\nn 
&&\quad\quad\quad+\bar H(r)^{-\frac{5}{8}}\left(-\tilde f(r) dt^2+dx_1^2+dx_2^2\right)\nn
e^{2\phi} & = & g_s^2\tilde H(r)^\half\\
\bar H(r) & = & 1+\frac{\rho_2^5\sinh^2\theta}{r^5}; \quad \tilde f(r) = 1-\frac{\rho_2^5}{r^5}\nonumber
\end{eqnarray}
It processes an well-defined temperature ($T\sim \frac{1}{\rho_2\cosh\theta}$) of the theory. However, for another set of values, $\d=-1$ and $\d_1=-\frac{4}{3}$, we get a deformed form of the black D$2$ brane which is given below.
\begin{eqnarray}
ds^2 & = & \bar H(r)^\frac{3}{8}\tilde f(r)^\frac{1}{40}\left(\frac{dr^2}{\tilde f(r)}+r^2d\Omega_6^2\right)\nn 
&&\quad\quad\quad+\bar H(r)^{-\frac{5}{8}}\tilde f(r)^{-\frac{1}{24}}\left(-\tilde f(r) dt^2+dx_1^2+dx_2^2\right)\nn
e^{2\phi} & = & g_s^2\tilde H(r)^\half\tilde f(r)^{-\half}
\end{eqnarray}
At the regime $\rho_2\ll r$, this can merge to the standard black D$2$ brane, assuming $\tilde f(r)^\frac{1}{40}\approx 1-\frac{1}{40}\frac{\rho_2^5}{r^5}\approx 1$ and $\tilde f(r)^{-\frac{1}{24}}\approx 1+\frac{1}{24}\frac{\rho_2^5}{r^5}\approx 1$. Therefore, in general the non-susy anisotropic D$2$ brane reduces to the black D$2$ brane at $\d=-1$. This is a particular case of the Hawking-Page transition which represents the deconfinement transition of the holographically dual gauge theory. In this respect, we can also conclude that the anisotropy parameter $\d$ is one of the controllers of the temperature.

\section{Decoupled geometry}
Same as the known AdS/CFT correspondence, the non-supersymmetric gauge/gravity duality \cite{Nayek:2015tta} also requires that the two theories on each side of the duality to be entirely decoupled from each other at the low energy limit. This needs the gravitational excitations in the bulk to be decoupled from the worldvolume gauge theory on the brane. At low energy scale, the nature of the scattering cross-sections for these excitations confirms the decoupling of the near brane regime from the ten dimensional bulk. Such decoupled geometry \cite{Nayek:2016hsi} eventually produces a non-AdS background dual to the non-supersymmetric YM-like gauge theory. Exactly similar picture is observed even for anisotropic non-susy D2 brane. In the low energy decoupling limit, the string length scale becomes so small, i.e., $\ell_s\rightarrow0$, that all other length scales in the theory become much higher than $\ell_s$. Again, in decoupling limit, the charge of the brane is very large resulting in large radius of curvature of transverse space in string unit, so that the supergravity approximation remains valid in the corresponding energy scale. Now, following the decoupled geometry of the non-susy D$3$ brane \cite{Nayek:2016hsi}, we can find the same by re-scaling the length scale $\rho$ accordingly as given below.
\bea
&&\rho=\alpha'u \quad\quad \rho_2=\a'u_2\nn
&& \cosh^2\theta = \frac{L}{u_2^5\a'^2}
\label{decoupling}
\eea
where $\alpha'=\ell_s^2$ and $L=3\pi^2g_{YM}^2N_c$ is a well-known `tHooft coupling of $2+1$ dimensional Yang-Mills theory, $g_{YM}^2$ and $N_c$ respectively being the coupling for $2+1$D pure YM theory and the number of the color charges in that theory \cite{Aharony:1999ti}. In the bulk theory, $N_c$ is the number of the anisotropic non-susy D$2$ brane and $g_\text{YM}^2$ is related to the string coupling $g_s$ as $g_\text{YM}^2=2g_s\alpha'^{-\half}$. As $g_s$ is dimensionless, the Yang-Mills coupling in $2+1$ dimension has dimension of mass. Here $u$ is the new radial coordinate in string unit with mass dimension $+1$ and hence defines the transformed energy scale of the decoupled theory. Whereas $u_2$ represents the aforementioned constant fixed point on the energy scale having the dimension of mass. In this decoupling limit the harmonic functions in (\ref{HF1}) and (\ref{HF2}) reduce to the following form.
\be
G(\rho)\to 1+\frac{u_2^5}{u^5}\equiv G(u);\quad\quad f(\rho)\to \frac{L}{u^5\a'^2}
\ee 
Then using the above scaling (\ref{decoupling}) in (\ref{Stringframe}) we achieve the decoupled or throat geometry of the anisotropic non-susy D$2$ brane.
\begin{eqnarray}
\frac{ds^2}{\a'} & = & \sqrt{\frac{L}{u^5}}G(u)^{-\frac{1}{20}+\frac{21}{80}\d_1-\frac{2}{5}\d}\left(\frac{du^2}{G(u)}+u^2d\Omega_6^2\right)\nonumber\\
&& +\sqrt{\frac{u^5}{L}}G(u)^{\frac{1}{4}+\frac{7}{16}\d_1-\d}\left(-G(u)^\d dt^2+dx_1^2+dx_2^2\right)\nn
e^{2\phi} &=& \frac{g_s^2}{\a'}\sqrt{\frac{L}{u^5}} G(u)^{-\frac{1}{4}+\frac{21}{16}\d_1-2\d}\nn
F_{[6]} & = & \hat{Q}\text{Vol}(\Omega_6)
\label{lowenergydecoupled}
\end{eqnarray} 
This geometry is quite clearly a non-AdS one without any conformal symmetry. The radius of curvature of the transverse six dimensional sphere is proportional to $L$. At the low energy limit as $\alpha'\to 0$, along with the planner limit $N_c\to\infty$, we take the perturbative gravity theory, i.e. $g_s\to 0$. This ensures a large finite value of $L$ and further the validity of the supergravity in the decoupled background. The dual gauge theory becomes strongly coupled because of the large $g_\text{YM}^2N_c$. In the decoupled gravity theory, besides the number of brane $N_c$ we have a dimensionfull parameter $u_2$ and a dimensionless parameter $\d$. However we have only one dimensionfull independent quantity in the $2+1$dim YM theory at zero temperature which is the gauge coupling. Therefore the parameters $u_2$ is related to the gauge coupling (`tHooft coupling) $g_\text{YM}^2N_c$. Using the standard relation  $e^\phi\approx \frac{\lambda^{\frac{5}{2}}}{N_c}$ we can define the dimensionless effective gauge coupling $\lambda^2$. The expression of the dilaton in the low energy decoupled geometry \eqref{lowenergydecoupled} gives the effective gauge coupling as
\begin{equation*}
    \lambda^2=\frac{1}{(6\pi^2)^{\frac{4}{5}}}\left(\frac{L}{u}\right)\left(1+\frac{u_2^5}{u^5}\right)^{-\frac{1}{10}+\frac{21}{40}\delta_1-\frac{4}{5}\delta}
\end{equation*}
where we use $g_\text{YM}^2=2g_s\a'^{-\half}$ and $L=3\pi^2g_{YM}^2N_c$. It shows that the $\lambda$ is a monotonically increasing function of the energy scale $u$. Such  behaviour of coupling is very much similar to the QCD-like theory and also with the 2+1D YM theory studied from the isotropic non-susy D2 brane \cite{Chakraborty:2020sty}. It also depends on the background parameters $\delta$ and $u_2$. At fixed $u$, $\lambda^2$ decreases with the increasing absolute value of the parameter $\d$. At $\d=-1.0$, $\lambda^2\sim \frac{1}{u}$ same as the BPS D$2$ brane. The suitable range of $u$ must be chosen accordingly so that the effective coupling remains not so large to discard the validity of supergravity and also not so small to make holographic interpretation inappropriate. For $u\gg u_2$, $\lambda\sim \frac{g_\text{YM}^2N_c}{u}$ which is exactly similar as the BPS D2 brane. The gauge theory in this range is super-YM theory. In IR regime, for $u\ll u_2$, the $\lambda$ diverges as 
\begin{equation}
    \lambda^2=\frac{L}{(6\pi^2)^{\frac{4}{5}}}\frac{u_2^{-\frac{1}{2}+\frac{21}{8}\delta_1-4\delta}}{u^{\frac{1}{2}+\frac{21}{8}\delta_1-4\delta}}
\end{equation}
So below a particular energy scale $u$, we need to uplift the non-susy D$2$ brane to the eleven dimensional non-susy M$2$ brane. Following the isotropic non-susy D$2$ brane \cite{Chakraborty:2020sty}, one can easily find the range of the energy scale $u$ where the gauge theory is strongly coupled non-supersymmetric YM theory and the gravity theory is fully described by the type-II supergravity. Below that particular range the gauge theory is described by the non-supersymmetric M-theory which has been discussed in detail in \cite{Aharony:1999ti,Chakraborty:2020sty}.

Again at the BPS limit, i.e. $u_2\to 0$, the decoupled geometry \eqref{lowenergydecoupled} reduces to the AdS$_4\times$S$^6$ with a conformal factor exactly same as the throat geometry of the  BPS D$2$ branes \cite{Aharony:1999ti}. On the other hand, the non-supersymmetric YM theory reduces to the supersymmetric Yang-Mills theory on the $2+1$ dimensional worldvolume. In this limit, the decoupled geometry do not contain any dimensionfull parameter like $u_2$. This makes the theory conformally symmetric. This is also known as the extremal limit.

\section{Confinement}
The breaking of superconformal symmetry in the gauge theory discussed above stems for the prediction of achieving the color confinement at the low energy limit in analogy with the QCD-like theories studied so far. The effective gauge coupling in the worldvolume theory of non-susy D2 brane is an energy scale-dependent quantity which is found to increase monotonically as we move towards the low energy regime. Thus in this regime, because of the strong enough self coupling between the gluons, the gauge theory lives in the confined phase similar as the low energy non-perturbative QCD. The popular way to prove the existence of such phase is to study the binding potential of a largely separated quark antiquark pair. However in $2+1$ dimensional YM theory, this potential is logarithmic in perturbative limit, but at the strong coupling regime it has the form of linear confinement $V(r)\sim \sigma r$, where $r$ represents the distance of separation between quark and antiquark and $\sigma$ is the proportionality constant known as the QCD flux-tube tension. Therefore, presence of the non-zero $\sigma$ ensures the confined phase of the theory.

\begin{figure}[h]
    \centering
    \includegraphics[width=8cm,height=5cm]{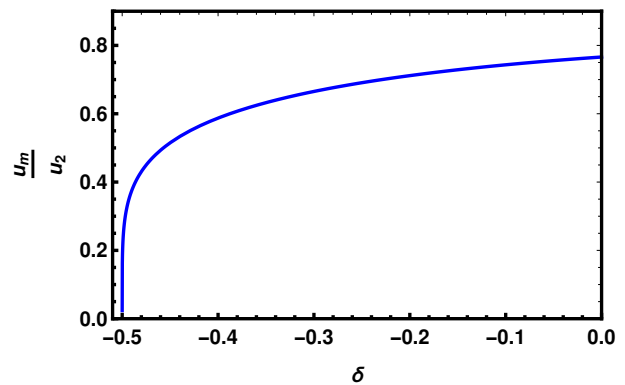}
    \caption{The plot of $\frac{u_m}{u_2}$ vs $\d$.}
    \label{um}
\end{figure}

In this section, our aim is to check whether there appears such confinement for the gauge theory living on the anisotropic non-susy D2 brane. To study this in holographic approach \cite{Constable:1999ch}, we consider a $Q-\bar{Q}$ pair on the $2+1$ dimensional boundary theory and an open string as a probe in the $3+1$ dimensional non-AdS decoupled geometry. Here the two ends of the probe string are connected with the $Q-\bar{Q}$ pair and rest of the string is hanging inside the bulk. Depending on the separation of the quark pair, the lowest point of the string moves along the radial direction $u$ in the bulk. Now it is convenient to find the binding potential of the quark pair using the thermal expectation value of the timelike Wilson loop with the relation, $W^F(\mathcal{C})=Exp\left[iS(\mathcal{C})\right]=Exp[iV(\mathcal{C})\tau]$ \cite{Maldacena:1998im,Rey:1998ik,Rey:1998bq,Brandhuber:1998er,Brandhuber:1998bs}. $\mathcal{C}$ is the timelike Wilson loop in gauge theory which is equivalent to the extremal worldsheet area swept out by the open string (probe) and $S(\mathcal{C})$ is the Nambo-Goto action calculated for that extremal worldsheet and $\tau$ defines the temporal length of the loop $\mathcal{C}$. In the decoupled geometry, Nambo-Goto action for the given open string is 
\begin{equation}
    S=\frac{1}{2\pi}\int d^2\zeta\sqrt{-\text{Det}\left(g_{\a\b}\right)}
    \label{NG}
\end{equation}
where the two dimensional pull-back metric is $g_{\alpha\beta}=G_{\mu\nu}\frac{\partial X^{\mu}}{\partial \zeta^{\alpha}}\frac{\partial X^{\nu}}{\partial \zeta^{\beta}}$, $X^{\mu}$ being the 10 dimensional bulk coordinates and $\zeta^{\alpha}, \alpha= 0, 1$ being the coordinates of the two-dimensional worldsheet drawn by the string. Here, the paramterization taken for the test string is 
\begin{equation}
\zeta^0=t, \zeta^1=x^1=x, u=u(x).
\label{parametrization}
\end{equation}
where x denotes the $Q-\bar{Q}$ separation. It is obvious from \eqref{parametrization} that the separation $x$ between $Q-\bar{Q}$ and the length of the test string depend on the energy scale of the decoupled theory. The worldsheet is taken to be localized at some constant values of other bulk coordinates. With such parametrization, the two dimensional metric components can be written as
\begin{eqnarray}
&& g_{\zeta^0\zeta^0}=G_{tt}\nn
&& g_{\zeta^0\zeta^1}=g_{\zeta^1\zeta^0}=0\nn
&& g_{\zeta^1\zeta^1}=G_{x^1x^1}+G_{uu}\left(\frac{\partial u}{\partial x}\right)^2
\end{eqnarray}
Substituting these we get the action \eqref{NG} as 
\begin{equation}
    S = \frac{1}{2\pi}\int dtdx P(u)\left[1+M^2(u)\left(\frac{d u}{d x}\right)^2\right]^{\frac{1}{2}}
\end{equation}
where, 
\begin{eqnarray}
    P(u) & = & \sqrt{\frac{u^5}{L}}G(u)^{\frac{1}{4}+\frac{7}{16}\d_1-\frac{1}{2}\d}\\
    M(u) & = & \sqrt{\frac{L}{u^5}}G(u)^{-\frac{13}{20}-\frac{7}{80}\d_1+\frac{3}{10}\d}
\end{eqnarray}
This action depends on the slope $\frac{du}{dx}$ defining the rate of change of the vertical distance of the string from the singularity inside the bulk with the length of separation $x$ between $Q-\bar{Q}$ pair. Now to get a stable configuration we need to set $\frac{du}{dx}=0$ and in that situation the worldsheet area is completely given by $P(u)$. So to find the minimal area we need to find the global minima of $P(u)$. We find that the global minima of $P(u)$ is located at 
\begin{equation}
    u_m=\left(\frac{7}{8}\d_1-\d-\frac{1}{2}\right)^\frac{1}{5}u_2
\end{equation}
for the geometry under consideration. The term within the parenthesis in the above expression of $u_m$ must be real and positive in order to achieve confinement. This is precisely maintained by modifying the valid range of $\d$ to $-\frac{1}{2}<\d\leq 0$. Again We have seen in the last section that $u$ can not be boundlessly small because of the validation of the supergravity theory. So at this point we need to be conscious so that the bottom of the string does not reach very close to the singularity. Plotting $u_m$ against $\d$ in Figure \ref{um}, we can see the string goes very close to the singularity near $\d=-0.5$. So, in view of the validation of the supergravity description, this particular method to find the confinement is allowed only when $\d$ is far away from $-0.5$.
Now at the minimum $u=u_m$, the binding potential of the $Q-\bar{Q}$ pair at the maximum separation $\Delta x$ between the quark and the antiquark is found to be
\bea
V & = & \frac{1}{2\pi}\int dx P(u_m)\nn
 & = & \frac{\Delta x}{2\pi}\sqrt{\frac{u_2^5}{8L}}\left(7\d_1-8\d-4\right)^{\frac{1}{4}-\frac{7}{16}\d_1+\frac{1}{2}\d}\nn
  &&\quad\quad\quad \times\left(7\d_1-8\d+4\right)^{\frac{1}{4}+\frac{7}{16}\d_1-\frac{1}{2}\d}
  \label{potential}
    \eea
Here $\Delta x=\int dx$ being the maximum separation of the quark pair is the maximum length of the QCD flux-tube. The binding potential is linearly proportional to the flux-tube length with a non-trivial proportionality constant. This constant coefficient of $\Delta x$ in the equation \eqref{potential} is identified as the tension of the flux tube $\sigma$.
\begin{eqnarray}
\sigma & = & \frac{1}{2\pi}\sqrt{\frac{u_2^5}{8L}}\left[(7\d_1-8\d)^2-16\right]^\frac{1}{4}\nn&&\quad\quad\times\left(\frac{7\d_1-8\d+4}{7\d_1-8\d-4}\right)^{\frac{7}{16}\d_1-\frac{1}{2}\d}
\label{sigma}
\end{eqnarray}
The QCD flux-tube tension $\sigma$ depends on the dimensionfull quantities $u_2$ and $L$ and dimensionless parameter $\d$. The mass dimension of $\sigma$ is $+2$. Like the non-susy isotropic D$2$ brane \cite{Chakraborty:2020sty}, we can set the fixed energy scale of the gravity background $u_2=g_{YM}^2N_c$ and can find the dimensionless quantity $\frac{\sqrt{\sigma}}{g_{YM}^2N_c}$ as,
\begin{eqnarray}
 \frac{\sqrt{\sigma}}{g_{YM}^2N_c} & = & \frac{1}{2\pi}\left(\frac{1}{6}\right)^{\frac{1}{4}}\left[(7\d_1-8\d)^2-16\right]^\frac{1}{8}\nn&&\quad\quad\times\left(\frac{7\d_1-8\d+4}{7\d_1-8\d-4}\right)^{\frac{7}{32}\d_1-\frac{1}{4}\d}
 \label{ssbgn}
\end{eqnarray}
For different values of $\delta$ within its specified range, the QCD string tension is tabulated below in Table \ref{sig}.
\begin{table}[h]
    \centering
    \caption{Values of $\sqrt{\sigma}$ at $u_2=g_\text{YM}^2N_c$ using \eqref{ssbgn}.}
    \begin{tabular}{|c|c|c|c|c|c|c|c|c|}
        \hline
        $\d$ & $0.0$ & $-0.1$ & $-0.2$ & $-0.3$ & $-0.4$ & $-0.49$ \\
        \hline
        $(-\d)^{1/5}$ & $0.0$ & $0.63$ & $0.72$ & $0.78$ & $0.83$ & $0.86$ \\ 
         \hline
         $\frac{\sqrt{\sigma}}{g_\text{YM}^2N_c}$ & $0.2010$ & $0.1980$ & $0.1942$ & $0.1891$ & $0.1824$ & $0.1729$ \\ 
         \hline
    \end{tabular}
    \label{sig}
\end{table}
Here, we can see that $\sigma$ is decreasing with the increasing values of $-\d$ and has maxima at $\d=0$. We have seen in the previous section that the effective gauge coupling $\lambda^2$ decreases with the increasing $-\d$. This indicates that the flux-tube tension $\sigma$ is proportional to the effective gauge coupling $\lambda^2$. As the background \eqref{lowenergydecoupled} reduces to the isotropic D$2$ brane at $\d=0$, the value of $\frac{\sqrt{\sigma}}{g_\text{YM}^2N_c}$ at $\d=0$ in the present solution matches with isotropic case \cite{Chakraborty:2020sty} with $\gamma=1$ (where $\gamma$ is defined in that article.). At this point it can be checked that at $\d=0$, $\frac{\sqrt{\sigma}}{g_\text{YM}^2N_c}$ get the value $0.2010$ which has good agreement with the lattice QCD \cite{Teper:1998te} and the theoretical prediction \cite{Nair:2002yg}. Since we have seen $\d=-1$ is supposed to represent the transition temperature of the confinement-deconfinement transition in the gauge theory, the flux-tube tension should vanish after crossing that point \cite{Bicudo:2017uyy}. However we are unable to compare this behaviour in our case as the method used herein does not allow us the whole parametric range.    

Now in the BPS limit \cite{Nayek:2016hsi}, $u_2\to 0$, $\sigma$ becomes zero according to the definition \eqref{sigma}. This is also consistent in another view. As we have seen earlier that at the BPS limit the considered non-susy anisotropic background reduces to the supersymmetric (BPS) D$2$ brane background. The BPS D$2$ brane processes the supersymmetric Yang-Mills theory as a dual theory in its worldvolume. The super-YM theory is known to be a deconfined theory, so $\sigma$ does not exist there.

\section{Mass spectrum of $0^{-+}$}
We know that, at the low energy (below a critical energy scale), the self-coupling of the gluons is much stronger which prevents the existence of the free gluons. These self-coupling gluons form the bound state, named as glueball. These bound states take a major role in the dynamics of the non-perturbative QCD. Eventually, the idea of glueball is entirely theoretical. Till date, there is no experimental evidence of these. According to the bag model \cite{Chodos:1974je,Chodos:1974pn,Jaffe:1975fd}, there are two fundamental modes of gluon field -- transverse electric (TE) mode with even parity and transverse magnetic (TM) mode with odd parity. The singlet bound state of these two different modes, (TETM), gives the pseudoscalar glueball $0^{-+}$ having zero spin, odd parity and even charge conjugation. The various different combinations of these modes give other glueballs like $0^{++},\,2^{++},\,2^{-+},\,1^{+-}$ etc (where $J$ of the form $J^{PC}$ indicates the spin of the bound state). In this section we will evaluate the mass spectrum of the pseudoscalar glueball which is one of the non-trivial properties of non-perturbative QCD$3$. The discreteness of the spectrum is shown theoretically \cite{Mathieu:2008me}. Some articles have calculated the numerical value of the mass in lattice QCD \cite{Philipsen:1996af} and also with holographic approaches \cite{Ooguri:1998hq,Csaki:1998qr}. Here to study the mass spectrum of $0^{-+}$ in $2+1$ dimensional QCD-like theory, in holographic model, we consider non-susy anisotropic D$2$ brane. According to the holographic approximation, this glueball mode is associated with the axion field of the bulk theory. So its mass can be found from the excitation of the bulk's axion field \cite{Ooguri:1998hq}. Although the decoupled gravity theory consists the vanishing axion field, it can have a non-trivial fluctuation $\chi$. The linearised equation of this fluctuation in the string frame is presented as
\begin{equation*}
    \frac{1}{\sqrt{-G}}\partial_\mu\left(\sqrt{-G}G^{\mu\nu}\partial_\nu\right)\chi=0
\end{equation*}
where $G_{\mu\nu}$ is the background metric \eqref{lowenergydecoupled} and $G$ is its determinant. We are interested in the most simplified low energy states of the glueball. So we consider the glueball with zero angular momentum i.e., the fluctuation has spherical symmetry in the six dimensional transverse sphere. Further $\chi$ is supposed to propagate along the worldvolume. Let us now take the axion fluctuation $\chi$ as 
\begin{equation*}
    \chi=f(u)e^{ik_ax^a}
\end{equation*}
where $k^a$ is the four-momentum $(E,\vec{p})$ and it gives the glueball mass as $M^2=E^2-p^2$. With the above ansatz, the equation of fluctuation of axion field becomes a second order differential equation.
\bea
&& \partial_u^2f+\left[\frac{7}{2u}+\left(\frac{3}{4}+\frac{21}{16}\delta_1-2\delta\right)\frac{\partial_u G}{G}\right]\partial_u f\nn
&&+\frac{L}{u^5}G(u)^{-\frac{13}{10}-\frac{7}{40}\delta_1+\frac{3}{5}\delta}\left[G(u)^{-\delta}E^2-p^2\right]f=0
\eea
We will now take a coordinate transformation $u=u_2e^{y}$. Because of this choice, the $0<u\leq u_2$ regime is zoomed into $-\infty<y\leq 0$ which simplifies our numerical analysis. Then we replace $f(y)$ as 
\begin{equation*}
    f(y)=e^{-\frac{5y}{4}}\left(1+e^{-5y}\right)^{-\frac{3}{8}-\frac{21}{32}\delta_1+\delta}\kappa(y)
\end{equation*}
After substitution and algebraic simplification we get the standard Schr\"odinger-like wave equation, 
\begin{equation}
    \kappa^{''}(y)-V(y)\kappa(y)=0
    \label{waveeq}
\end{equation}
with the potential function,
\bea
V(y) & = & \frac{25}{16}+\frac{25}{4}\frac{\left(\frac{3}{4}+\frac{21}{16}\delta_1-2\delta\right)}{1+e^{5y}}+\frac{25}{4}\frac{\left[\left(\frac{1}{4}-\frac{21}{16}\delta_1+2\delta\right)^2-1\right]}{\left(1+e^{5y}\right)^2}\nn
 &&\quad\quad -e^{-3y}(1+e^{-5y})^{-\frac{13}{10}-\frac{7}{40}\delta_1-\frac{2}{5}\delta}\omega^2
 \label{wavepot}
\eea
where $m^2=\frac{LM^2}{u_2^3}$ and $\omega^2=\frac{LE^2}{u_2^3}$ and assuming the linear momentum of glueball is negligible in confined state, i.e., $E\gg p$. Now to understand the detail nature of this potential function we first consider its asymptotic behaviour. At the positive and negative infinity of $y$, the potential \eqref{wavepot} takes the following form.
\bea
 V(y)|_{y\rightarrow+\infty} & = & \frac{25}{16}-e^{-3y}m^2\nonumber\\
 V(y)|_{y\rightarrow-\infty} & = & \frac{25}{4}\left(\frac{1}{4}+\frac{21}{16}\delta_1-2\delta\right)^2-e^{\left(\frac{7}{2}+\frac{7}{8}\delta_1+2\delta\right)y}\omega^2\nonumber
\eea
 So at positive asymptote the potential merges to constant value $\frac{25}{16}$. As we now decrease $y$ we find the point $y_+$ where $V$ changes sign. It behaves in similar manner at the negative asymptote where it merges to the positive constant $\frac{25}{4}\left(\frac{1}{4}+\frac{21}{16}\delta_1-2\delta\right)^2$ and changes sign at $y=y_-$. So one can easily visualize that the potential forms a small well of negative potential with the boundary $y_-\leq y\leq y_+$. The associated turning points of the potential-well are
\bea
y_+ & = & \frac{2}{3}\ln\left[\frac{4}{5}m\right]\nn
y_- & = & \frac{12}{21+20\delta+\sqrt{21-20\d^2}}\ln\left[\frac{5}{8}\frac{1+\sqrt{21-20\d^2}}{\omega}\right]\nonumber
\eea
Now the concept is as follows : the function $\kappa(y)$ in \eqref{waveeq} corresponding to the pseudoscalar glueball $0^{-+}$ is a normalized wave function of the potential-well \eqref{wavepot}. Therefore, assuming the depth of the well is small enough, the potential follows the WKB quantization condition as given below.
\begin{equation}
    \int_{y_-}^{y_+}dy\sqrt{-V(y)}=\left(n+\frac{1}{2}\right)\pi
    \label{wkb}
\end{equation}
The last term of the potential \eqref{wavepot}, which is directly proportional to $\omega^2$ ($\approx m^2$), controls the depth of the potential-well. So the WKB approximation can be used for the low energy states only. In \eqref{wkb}, $n$ indicates the label of various excited states of the pseudoscalar glueball, e.g., the ground state ($n=0$), the first excited state ($n=1$), the second excited state ($n=2$) and so on. So the above approximation is applicable for the first few $n$ values. The left hand side of \eqref{wkb} is a function of $m$ and $\d$ where the right hand side is a constant for a given level $n$. So evaluating the integration in \eqref{wkb}, we can find the mass $m$ for given $\d$ and $n$. 

The complicated form of the potential $\eqref{wavepot}$ makes it harder to solve the WKB equation \eqref{wkb} analytically. However, following \cite{Nayek:2021ded}, one can find some analytical expressions of $m$ or $\omega$ by using the Taylor series expansion of the integrand in \eqref{wkb} assuming large $\omega$. Otherwise $m$ can be solved using numerical technique, which is going to be used in this article. In the numerical technique, for given $\d$ and $n$, we first find the boundary of the potential-well ($y_-,\,y_+$) from the roots of the equation $V(y)=0$. Next, using the shooting method we find the value of $\omega$ which satisfies the equality condition of \eqref{wkb}. Here the Table \ref{tab:mass} has listed the values of $m$ for few sets of ($n,\,\d$) using the numerical technique.
\begin{table}[t]
    \centering
    \caption{The mass $m$ has been listed for different values of $\d$ for the ground state ($n=0$), first excited state ($n=1$) and second excited state ($n=2$). The mass ratio of the first excited state and ground state has been presented in the fifth column. In the sixth column, the mass-to-tension ratio has been shown for the ground state.}
    \begin{tabular}{c|c|c|c|c|c}
    \hline\hline
    $\d$ & $n=0$ & $n=1$ & $n=2$ & $\frac{m_{0^{-+}}^*}{m_{0^{-+}}}$& $\frac{M_{0^{-+}}}{\sqrt{\sigma}}$\\
    \hline
    $0.0$ & $5.19162$ & $8.49143$ & $11.7237$ & $1.6356$ & $4.745$\\
    $-0.1$ & $5.11806$ & $8.29351$ & $11.3957$ & $1.62044$ & $4.748$\\
    $-0.2$ & $5.02796$ & $8.06344$ & $11.0218$ & $1.60372$ & $4.757$\\
    $-0.3$ & $4.91824$ & $7.79513$ & $10.5931$ & $1.58494$ & $4.777$\\
    $-0.4$ & $4.78409$ & $7.47973$ & $10.0977$ & $1.56346$ & $4.819$\\
    $-0.5$ & $4.61774$ & $7.10415$ & $9.518$ & $1.53845$ & \\
    $-0.6$ & $4.4062$ & $6.64836$ & $8.82754$ & $1.50886$ & \\ 
    $-0.7$ & $4.1262$ & $6.0799$ & $7.98421$ & $1.47349$ & \\
    $-0.8$ & $3.73338$ & $5.34067$ & $6.91322$ & $1.43052$ & \\
    $-0.9$ & $3.13331$ & $4.30372$ & $5.45271$ & $1.37354$ & \\
    $-1.0$ & $2.07339$ & $2.46851$ & $2.98422$ & $1.19057$ & \\
    \hline\hline
    \end{tabular}
    \label{tab:mass}
\end{table}
We find pseudoscalar glueball mass in the unit of $\sqrt{u_2^3/L}$, i.e., $M=m\sqrt{u_2^3/L}$. Again, we know that, for the non-perturbative QCD-like gauge theory, $u_2\sim g_\text{YM}^2N_c$ which finally expresses the mass in the unit of $g_\text{YM}^2N_c$. At $\d=0$, the states are found to have maximum masses. So for the isotropic gravity background the glueball has largest mass and it decreases as the anisotropy is introduced. Due to the increase of the anisotropry parameter, $-\d$, the energy gap between two energy states also decreases. At $\d=-1$, the masses of the first few energy levels take small but finite values and are also found to differ from each other with small gap. This behaviour of the spectrum is also presented pictorially.
\begin{figure}[h]
    \centering
    \includegraphics[width=8cm,height=5cm]{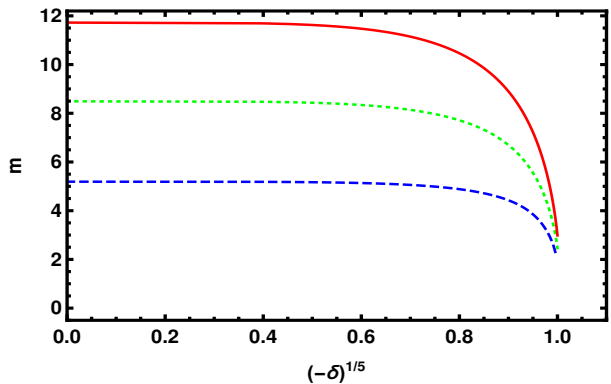}
    \caption{The plot of $m$ vs $(-\d)^{1/5}$ for various energy levels; $n=0$ (blue), $n=1$ (green) and $n=2$ (red).}
    \label{md}
\end{figure}
In the Figure \ref{md}, the pseudoscalar glueball mass $m$ have been plotted against the parameter $(-\d)^{1/5}$ for first three energy levels. The masses have been found to maintain almost a constant value up to $(-\d)^{1/5}\approx 0.7$ and shown an abrupt variation as it goes very close to the transition point $(-\d)^{1/5}\approx 1$. These are similar to the study of the anisotropic non-susy D$3$ brane \cite{Nayek:2021ded} where the same has been studied in $3+1$ dimensional QCD-like theory. Therefore, from the holographic study of the $2+1$ dimensional QCD-like theory under the deconfinement transition, we can have two important observations as follow. Firstly, for a given energy level, the mass significantly decreases as the parameter $\d$ moves towards the Hawking-Page transition point. It is consistent with our prediction since we know this transition gives the deconfinement transition in the dual QCD-like gauge theory and the glueballs are supposed to exist only in the confined phase. Second one is the contraction of the mass gap between two energy levels as they head towards the transition point. The mass can depend on $\d$ either through the effective gauge coupling $\lambda^2$ or through the explicit presence of $\d$ in $m$. We have seen that $\lambda^2$, for fixed energy scale, decreases with the increasing value of $-\d$. As the effective gauge coupling becomes weaker the glueball mass is expected to decrease. So our first observation can be explained easily referring the $\lambda^2$-dependence of $m$. In \cite{Chakraborty:2020sty}, the variation of the pseudoscalar glueball mass has been studied with respect to the parameter which represents the gluon condensation at zero temperature. There the mass gap between two energy levels remains constant however the mass of the individual state varies (TABLE II in \cite{Chakraborty:2020sty}). So to argue the second observation, we expect that the mass of the glueball explicitly depends on the parameter $\d$. 
\begin{figure}[t]
    \centering
    \includegraphics[width=8cm,height=5cm]{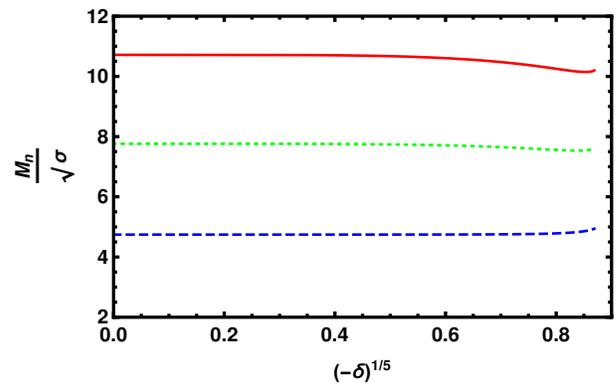}
    \caption{The plot of $\frac{M_n}{\sqrt{\sigma}}$ vs $(-\d)^{1/5}$ where $M_n$ is the mass of the $n^\text{th}$ excited state of the pseudoscalar glueball; $n=0$ (blue), $n=1$ (green) and $n=2$ (red).}
    \label{ms}
\end{figure}
Now like the case of the flux-tube tension if we take $u_2=g_\text{YM}^2N_c$, the dimensionfull mass $M$ of the glueball can be written in terms of $m$ as $M=\frac{m}{\sqrt{3}\pi}g_\text{YM}^2N_c$ which is in the same unit as $\sqrt{\sigma}$. In Table \ref{tab:mass}, we have seen that the ratio $\frac{M_{0^{-+}}}{\sqrt{\sigma}}$ is almost constant for the ground state. However we have discussed before, due to the validity of supergravity approximation, we are not able to calculate $\sigma$ beyond a certain value of $\d$. The ratio has been also plotted in Figure \ref{ms} for the ground state, first excited state and the second excited state. In all of these three cases $\frac{M_n}{\sqrt{\sigma}}$ have been found to have the constant value, where $M_n$ indicates the mass of the $n$-th energy state. Like the isotropic case \cite{Chakraborty:2020sty}, the ratio differs with the factor of $3n$ for the various states, i.e., the energy gap between two states $M_{n_1}-M_{n_2}=3(n_1-n_2)\sqrt{\sigma}$.

\section{Conclusion}
The behavior of the $2+1$ dimensional nonperturbative quantum chromodynamics is known to have various qualitative similarities with the $3+1$ dimension. Here we have presented the same considering two particular quantities of QCD -- flux-tube tension $\sigma$ and the pseudoscalar glueball spectrum $M$. The anisotropic non-susy D$2$ brane has been considered as the holographic dilatonic gravity model. This brane solution is non-BPS. Generally, this model is supposed to contain the open string tachyon mode in the worldvolume theory for finite charge of the brane and the gauge theory is the Yang-Mills theory in presence of the tachyonic mode. But, in the decoupling limit, the charge of the brane has been taken to be infinitely large which has suppressed the tachyonic mode from the gauge theory. It has reduced the gauge theory to the Yang-Mills theory with some non-trivial excitations without any tachyon \cite{Lu:2007bu,Sen:1999mg,Sen:2004nf}. In this article, the non-AdS decoupled background \eqref{lowenergydecoupled} has indicated the non-trivial fixed point on the energy scale and the dilatonic profile has given the energy scale dependent gauge coupling. So the corresponding gauge theory is expected to be similar with the QCD in $2+1$ dimensions. Gravity theory contains the parameters $u_2$ and $\d$ which are supposed to be related to the dimensionfull `t Hooft coupling ($g_\text{YM}^2N_c$) and temperature $T$ of that $2+1$ dimensional QCD-like gauge theory. At $\d=-1$, the gravity geometry has been found to reduce to the black D$2$ brane with temperature $T_c\sim u_2$, i.e., the Hawking-Page transition. Also the gravity background asymptotically reduces to the form AdS$_4\times$S$^6$. Therefore, following the Hawking-Page transition of the non-susy anisotropic D$3$ brane \cite{Nayek:2021ded}, we can assume an empirical definition of the temperature as 
\begin{equation}
    T=\left(-\d\right)^{1/5}T_c.
\end{equation}      
So, varying the parameter $\d$, we can go from the confined theory to the deconfined theory, at $T=T_c$. Therefore the corresponding gauge theory is the $2+1$ dimesional Yang-Mills theory with some excitations at the finite temperature $T$. In this article we have found the variation of the QCD flux-tube tension $\sigma$ with $T$ up to $T/T_c\approx 0.8$. We have noticed that the flux-tube tension has been decreased with $T$ although the slope is very small which can be backed with the following fact. Here the method we have used to show the confinement did not allow us to access the high temperature regime due to the violation of the supergravity approximation. So we have observed the confinement in the low temperature regime only and in that regime, as the theory is completely confined, we have not observed any non-trivial change of $\sigma$ with $T$ in Table \ref{sig}. Considering the fluctuation of the axion field, we have shown the variation of the pseudoscalar glueball mass with $\d$ in Table \ref{tab:mass}. The explicit $\d$-dependence of the spectrum indicates that the mass of a given energy state depends on $T$ along with the effective gauge coupling $\lambda^2$. In Figure \ref{md} we have plotted the variation of mass with $T/T_c=(-\d)^{1/5}$ for first three lowest energy levels. $M$'s have maximum value at $T=0$. As we have increased $T$, $M$ has been found to decreased slowly for the lower value of $T$ and near $T/T_c=1$ they have shown a significant change. We have also seen that mass gap of two consecutive energy levels of the glueball is directly related to $T$. At $T=T_c$ this gap has been reduced to a trivial order of the mass. 

In a recent article \cite{Brandt:2017yzw}, with the higher order corrections in the effective string action, the value of the flux-tube tension has been predicted to be $\sqrt{\sigma}=486.4$MeV for SU($3$) and $\sqrt{\sigma}=487.6$MeV for SU($2$) at zero temperature. Using our definition \eqref{ssbgn} with $\d=0$, these indicate $g_\text{YM}^2N_c$ to be equal to $2419.9$MeV and $2425.87$MeV respectively. With these data, our holographic results given in Table \ref{tab:mass} predicts the ground state mass of $0^{-+}$ is to be $2.3$GeV at $T=0$ and $0.9$GeV at $T=T_c$, where for the first excited state these are $3.8$GeV and $1.1$GeV respectively. The mass gap between these two state has been decreased from $1.5$GeV to $0.2$GeV as temperature changes from $0$ to $T_c$, which is also a non-trivial change of spectrum. At $T=T_c$ the non-susy anisotropic D$2$ brane has been transformed in to the black D$2$ brane. Holographically, at this particular temperature $T_c$, the $2+1$ dimensional QCD-like theory is expected to show the well-known QCD phase transition from the confinement to deconfinement\footnote{Some articles \cite{Mandal:2011ws,Bergner:2021goh} argue that the Hawking-Page transition brings the theory to a partial deconfinement. The complete deconfinement occurs at the point of the Gregory-Laflamme transition. }. In the deconfined phase the glueballs are supposed to be replaced by the free gluons. So our observed decay of the pseudoscalar glueball spectrum can be taken as signature of the deconfinement transition in the aforementioned QCD-like gauge theory.

Although we have presented some significant observations of $2+1$ dimensional Yang-Mills. Our study has been restricted with few constraints. Firstly, here we don't have the exact relation of $T$ with $\d$ and $u_2$ which can be found by studying the thermodynamics in this model. Secondly, as we have mentioned earlier that due to the validity of the supergravity approximation, our used method was unable to study the confinement for the whole confined phase, so to strengthen the observation of the $2+1$ dimensional QCD phase transition in this background we need to apply some other approach. Again, one can further study the other modes of glueball in this gravity background to check the existence of the transition at the concluded point. However the non-zero mass of the glueball at $T=T_c$ can be argued with the concept of the partial deconfinement, it will be interesting to study the same in the black D$2$ brane background. Apart from these, there are various properties of the non-perturbative QCD$3$ which are analogous to the QCD$4$, they should be studied in the present gravity model to explore the detail of the dual gauge theory.


\begin{thebibliography}{99}


\bibitem{Maldacena:1997re}
J.~M.~Maldacena,
Adv. Theor. Math. Phys. \textbf{2}, 231-252 (1998)
doi:10.1023/A:1026654312961
[arXiv:hep-th/9711200 [hep-th]].

\bibitem{Aharony:1999ti}
O.~Aharony, S.~S.~Gubser, J.~M.~Maldacena, H.~Ooguri and Y.~Oz,
Phys. Rept. \textbf{323}, 183-386 (2000)
doi:10.1016/S0370-1573(99)00083-6
[arXiv:hep-th/9905111 [hep-th]].

\bibitem{Nayek:2015tta}
K.~Nayek and S.~Roy,
JHEP \textbf{03}, 102 (2016)
doi:10.1007/JHEP03(2016)102
[arXiv:1506.08583 [hep-th]].

\bibitem{Nayek:2016hsi}
K.~Nayek and S.~Roy,
Phys. Lett. B \textbf{766}, 192-195 (2017)
doi:10.1016/j.physletb.2017.01.007
[arXiv:1608.05036 [hep-th]].

\bibitem{Lu:2004ms}
J.~X.~Lu and S.~Roy,
JHEP \textbf{02}, 001 (2005)
doi:10.1088/1126-6708/2005/02/001
[arXiv:hep-th/0408242 [hep-th]].

\bibitem{Witten:1998zw}
E.~Witten,
Adv. Theor. Math. Phys. \textbf{2}, 505-532 (1998)
doi:10.4310/ATMP.1998.v2.n3.a3
[arXiv:hep-th/9803131 [hep-th]].

\bibitem{Constable:1999ch}
N.~R.~Constable and R.~C.~Myers,
JHEP \textbf{11}, 020 (1999)
doi:10.1088/1126-6708/1999/11/020
[arXiv:hep-th/9905081 [hep-th]].

\bibitem{Ooguri:1998hq}
H.~Ooguri, H.~Robins and J.~Tannenhauser,
Phys. Lett. B \textbf{437}, 77-81 (1998)
doi:10.1016/S0370-2693(98)00877-6
[arXiv:hep-th/9806171 [hep-th]].

\bibitem{Csaki:1998qr}
C.~Csaki, H.~Ooguri, Y.~Oz and J.~Terning,
JHEP \textbf{01}, 017 (1999)
doi:10.1088/1126-6708/1999/01/017
[arXiv:hep-th/9806021 [hep-th]].

\bibitem{Babington:2003vm}
J.~Babington, J.~Erdmenger, N.~J.~Evans, Z.~Guralnik and I.~Kirsch,
Phys. Rev. D \textbf{69}, 066007 (2004)
doi:10.1103/PhysRevD.69.066007
[arXiv:hep-th/0306018 [hep-th]].

\bibitem{Csaki:2006ji}
C.~Csaki and M.~Reece,
JHEP \textbf{05}, 062 (2007)
doi:10.1088/1126-6708/2007/05/062
[arXiv:hep-ph/0608266 [hep-ph]].

\bibitem{Kim:2007qk}
Y.~Kim, B.~H.~Lee, C.~Park and S.~J.~Sin,
JHEP \textbf{09}, 105 (2007)
doi:10.1088/1126-6708/2007/09/105
[arXiv:hep-th/0702131 [hep-th]].

\bibitem{Polchinski:2002jw}
J.~Polchinski and M.~J.~Strassler,
JHEP \textbf{05}, 012 (2003)
doi:10.1088/1126-6708/2003/05/012
[arXiv:hep-th/0209211 [hep-th]].

\bibitem{Chakraborty:2017wdh}
S.~Chakraborty, K.~Nayek and S.~Roy,
Nucl. Phys. B \textbf{937}, 196-213 (2018)
doi:10.1016/j.nuclphysb.2018.10.010
[arXiv:1710.08631 [hep-th]].

\bibitem{Polchinski:2001tt}
J.~Polchinski and M.~J.~Strassler,
Phys. Rev. Lett. \textbf{88}, 031601 (2002)
doi:10.1103/PhysRevLett.88.031601
[arXiv:hep-th/0109174 [hep-th]].

\bibitem{Morningstar:1999rf}
C.~J.~Morningstar and M.~J.~Peardon,
Phys. Rev. D \textbf{60}, 034509 (1999)
doi:10.1103/PhysRevD.60.034509
[arXiv:hep-lat/9901004 [hep-lat]].

\bibitem{Teper:1997tq}
M.~Teper,
Phys. Lett. B \textbf{397}, 223-228 (1997)
doi:10.1016/S0370-2693(97)00181-0
[arXiv:hep-lat/9701003 [hep-lat]].

\bibitem{Miller:2006hr}
D.~E.~Miller,
Phys. Rept. \textbf{443}, 55-96 (2007)
doi:10.1016/j.physrep.2007.02.012
[arXiv:hep-ph/0608234 [hep-ph]].

\bibitem{Lucini:2001ej}
B.~Lucini and M.~Teper,
JHEP \textbf{06}, 050 (2001)
doi:10.1088/1126-6708/2001/06/050
[arXiv:hep-lat/0103027 [hep-lat]].

\bibitem{Brandt:2017yzw}
B.~B.~Brandt,
JHEP \textbf{07}, 008 (2017)
doi:10.1007/JHEP07(2017)008
[arXiv:1705.03828 [hep-lat]].

\bibitem{Mandal:2011ws}
G.~Mandal and T.~Morita,
JHEP \textbf{09}, 073 (2011)
doi:10.1007/JHEP09(2011)073
[arXiv:1107.4048 [hep-th]].

\bibitem{Bergner:2021goh}
G.~Bergner \textit{et al.} [MCSMC],
[arXiv:2110.01312 [hep-th]].

\bibitem{Horowitz:1991cd}
G.~T.~Horowitz and A.~Strominger,
Nucl. Phys. B \textbf{360}, 197-209 (1991)
doi:10.1016/0550-3213(91)90440-9

\bibitem{Maldacena:1998im}
J.~M.~Maldacena,
Phys. Rev. Lett. \textbf{80}, 4859-4862 (1998)
doi:10.1103/PhysRevLett.80.4859
[arXiv:hep-th/9803002 [hep-th]].

\bibitem{Rey:1998ik}
S.~J.~Rey and J.~T.~Yee,
Eur. Phys. J. C \textbf{22}, 379-394 (2001)
doi:10.1007/s100520100799
[arXiv:hep-th/9803001 [hep-th]].

\bibitem{Rey:1998bq}
S.~J.~Rey, S.~Theisen and J.~T.~Yee,
Nucl. Phys. B \textbf{527}, 171-186 (1998)
doi:10.1016/S0550-3213(98)00471-4
[arXiv:hep-th/9803135 [hep-th]].

\bibitem{Brandhuber:1998er}
A.~Brandhuber, N.~Itzhaki, J.~Sonnenschein and S.~Yankielowicz,
JHEP \textbf{06}, 001 (1998)
doi:10.1088/1126-6708/1998/06/001
[arXiv:hep-th/9803263 [hep-th]].

\bibitem{Brandhuber:1998bs}
A.~Brandhuber, N.~Itzhaki, J.~Sonnenschein and S.~Yankielowicz,
Phys. Lett. B \textbf{434}, 36-40 (1998)
doi:10.1016/S0370-2693(98)00730-8
[arXiv:hep-th/9803137 [hep-th]].

\bibitem{Chakraborty:2020sty}
A.~Chakraborty and K.~Nayek,
Phys. Rev. D \textbf{103}, 066001 (2021)
doi:10.1103/PhysRevD.103.066001
[arXiv:2008.00770 [hep-th]].

\bibitem{Karabali:1995ps}
D.~Karabali and V.~P.~Nair,
Nucl. Phys. B \textbf{464}, 135-152 (1996)
doi:10.1016/0550-3213(96)00034-X
[arXiv:hep-th/9510157 [hep-th]].

\bibitem{Karabali:1996je}
D.~Karabali and V.~P.~Nair,
Phys. Lett. B \textbf{379}, 141-147 (1996)
doi:10.1016/0370-2693(96)00422-4
[arXiv:hep-th/9602155 [hep-th]].

\bibitem{Karabali:1996iu}
D.~Karabali and V.~P.~Nair,
Int. J. Mod. Phys. A \textbf{12}, 1161-1172 (1997)
doi:10.1142/S0217751X9700089X
[arXiv:hep-th/9610002 [hep-th]].

\bibitem{Karabali:1998yq}
D.~Karabali, C.~j.~Kim and V.~P.~Nair,
Phys. Lett. B \textbf{434}, 103-109 (1998)
doi:10.1016/S0370-2693(98)00751-5
[arXiv:hep-th/9804132 [hep-th]].

\bibitem{Nair:2002yg}
V.~P.~Nair,
Nucl. Phys. B Proc. Suppl. \textbf{108}, 194-200 (2002)
doi:10.1016/S0920-5632(02)01328-2
[arXiv:hep-th/0204063 [hep-th]].

\bibitem{Hong:2010sb}
D.~K.~Hong and H.~U.~Yee,
JHEP \textbf{05}, 036 (2010)
[erratum: JHEP \textbf{08}, 120 (2010)]
doi:10.1007/JHEP05(2010)036
[arXiv:1003.1306 [hep-th]].

\bibitem{Teper:1998te}
M.~J.~Teper,
Phys. Rev. D \textbf{59}, 014512 (1999)
doi:10.1103/PhysRevD.59.014512
[arXiv:hep-lat/9804008 [hep-lat]].

\bibitem{Teper:1993gm}
M.~Teper,
Phys. Lett. B \textbf{311}, 223-229 (1993)
doi:10.1016/0370-2693(93)90559-Z

\bibitem{Philipsen:1996af}
O.~Philipsen, M.~Teper and H.~Wittig,
Nucl. Phys. B \textbf{469}, 445-472 (1996)
doi:10.1016/0550-3213(96)00156-3
[arXiv:hep-lat/9602006 [hep-lat]].

\bibitem{Lu:2007bu}
J.~X.~Lu, S.~Roy, Z.~L.~Wang and R.~J.~Wu,
Nucl. Phys. B \textbf{813}, 259-282 (2009)
doi:10.1016/j.nuclphysb.2009.01.005
[arXiv:0710.5233 [hep-th]].

\bibitem{Nayek:2021ded}
K.~Nayek and S.~Roy,
[arXiv:2105.01503 [hep-th]].

\bibitem{Hawking:1982dh}
S.~W.~Hawking and D.~N.~Page,
Commun. Math. Phys. \textbf{87}, 577 (1983)
doi:10.1007/BF01208266

\bibitem{Bicudo:2017uyy}
P.~Bicudo, N.~Cardoso and M.~Cardoso,
Nucl. Phys. B \textbf{940}, 88-112 (2019)
doi:10.1016/j.nuclphysb.2019.01.012
[arXiv:1702.03454 [hep-lat]].

\bibitem{Chodos:1974je}
A.~Chodos, R.~L.~Jaffe, K.~Johnson, C.~B.~Thorn and V.~F.~Weisskopf,
Phys. Rev. D \textbf{9}, 3471-3495 (1974)
doi:10.1103/PhysRevD.9.3471

\bibitem{Chodos:1974pn}
A.~Chodos, R.~L.~Jaffe, K.~Johnson and C.~B.~Thorn,
Phys. Rev. D \textbf{10}, 2599 (1974)
doi:10.1103/PhysRevD.10.2599

\bibitem{Jaffe:1975fd}
R.~L.~Jaffe and K.~Johnson,
Phys. Lett. B \textbf{60}, 201-204 (1976)
doi:10.1016/0370-2693(76)90423-8

\bibitem{Mathieu:2008me}
V.~Mathieu, N.~Kochelev and V.~Vento,
Int. J. Mod. Phys. E \textbf{18}, 1-49 (2009)
doi:10.1142/S0218301309012124
[arXiv:0810.4453 [hep-ph]].

\bibitem{Sen:1999mg}
A.~Sen,
[arXiv:hep-th/9904207 [hep-th]].

\bibitem{Sen:2004nf}
A.~Sen,
Int. J. Mod. Phys. A \textbf{20}, 5513-5656 (2005)
doi:10.1142/S0217751X0502519X
[arXiv:hep-th/0410103 [hep-th]].

\end{thebibliography}
\end{document}